\documentclass{PoS}
\usepackage{slashed,amsmath,amssymb}

\title{Schwinger Model Mass Anomalous Dimension}

\ShortTitle{Schwinger Model Mass Anomalous Dimension}

 \author{\speaker{Liam Keegan}
 \thanks{While the talk was cancelled due to illness of the presenter,
         this contribution is made available by discretion of the editorial
         board of the LATTICE 2015 proceedings.} \\
 \\ PH-TH, CERN, CH-1211 Geneva 23, Switzerland\\
         E-mail: \email{liam.keegan@cern.ch}}

\abstract{The mass anomalous dimension for several gauge theories with an infrared fixed point has recently been determined using the mode number of the Dirac operator. In order to better understand the sources of systematic error in this method, we apply it to a simpler model, the massive Schwinger model with two flavours of fermions, where analytical results are available for comparison with the lattice data.}

\FullConference{The 33rd International Symposium on Lattice Field Theory\\
                 14-18 July, 2015\\
                 Kobe International Conference Center, Kobe, Japan}

\begin{document}

\section{Introduction}
Recently the mode number of the Dirac operator on the lattice has been used to determine the mass anomalous dimension for both confining and conformal theories~\cite{DeGrand:2009hu,DelDebbio:2010ze,Cheng:2011ic,Patella:2011jr,Hasenfratz:2012fp,Patella:2012da,deForcrand:2012vh,Cheng:2013eu,Cheng:2013bca,Landa-Marban:2013oia,Cichy:2013eoa,DelDebbio:2013hha,Perez:2015yna}. This method has been shown to work well in the case of confining theories, where the running of the mass anomalous dimension can be followed from the perturbative regime through to the confining regime by fitting to different ranges of eigenvalues and combining results from several lattice spacings~\cite{Cheng:2013eu}. For QCD this has recently been done with the addition of a continuum extrapolation, which gives very good agreement with perturbation theory~\cite{Cichy:2013eoa}. For theories with an IRFP the method gives very precise values, see e.g. Ref.~\cite{Patella:2012da}, although the dependence on the fit range and bare coupling is less clear. 
As stated in Ref.~\cite{Landa-Marban:2013oia} it would clearly be desirable to apply the method to a theory where analytic results are available for comparison.	

In this work we apply the mode number method to the $n_f=2$ massive Schwinger model. The Schwinger model has a long history as a toy model for lattice simulations, see e.g~\cite{Gattringer:1997qc,Gutsfeld:1999pu,Gattringer:1999gt,Giusti:2001cn,Christian:2005yp,Bietenholz:2011ey} and references therein. This is a very attractive toy model for testing the mode number method, as in addition to having an analytic solution for the mass anomalous dimension, it lives in only two dimensions and so is inexpensive to simulate numerically. This was recently done in Ref.~\cite{Landa-Marban:2013oia}, but only a single gauge coupling was considered and the expected value for $\gamma_*$ was not found. Here we show that by measuring $\gamma$ as a function of the eigenvalue and using several different bare couplings, as advocated in Ref.~\cite{Cheng:2013eu}, it is possible to follow the running of the mass anomalous dimension over a range of energy scales from the UV to the IR, where we find a value which approaches the analytic result. We also apply the method to the $n_f=0$ theory which has a non--zero chiral condensate, where we find the expected behaviour in the IR.

\section{Mode Number Method}
In a mass--deformed conformal field theory (mCFT) in $d$ dimensions, the chiral condensate goes to zero with the mass as
\begin{equation}
\label{eq:condensate}
\left\langle \overline{\psi}\psi \right\rangle \propto m^{\frac{d}{1+\gamma_*} - 1},
\end{equation}
where $\gamma_*$ is the mass anomalous dimension at the IRFP.

The spectral density $\rho(\omega)$ of the Dirac operator at small eigenvalues $\omega$ also goes to zero with the same exponent~\cite{DeGrand:2009hu,DelDebbio:2010ze} 
\begin{equation}
\label{eq:rho}
\lim_{m\rightarrow 0}\lim_{V\rightarrow\infty}\rho(\omega) \propto \omega^{\frac{d}{1+\gamma_*} - 1}.
\end{equation}

In addition, for an asymptotically free theory in the UV limit this equation can be written in the same form but with $\gamma_*$ replaced by $\gamma(g^2_R)$, the 1--loop mass anomalous dimension which is a function of the renormalised coupling $g^2_R$~\cite{Cheng:2013eu},
\begin{equation}
\lim_{m\rightarrow 0}\lim_{V\rightarrow\infty}\rho(\omega) \propto \omega^{\frac{d}{1+\gamma(g^2_R)} - 1}.
\end{equation}

The idea is then to fit the measured spectral density to an equation of this form to extract an effective mass anomalous dimension $\gamma(\lambda)$ as a function of the eigenvalue $\lambda$. This quantity is equal to $\gamma_*$ in the IR limit $\lambda\rightarrow0$, and to the perturbative 1--loop $\gamma(g^2_R)$ in the UV limit.

In a theory with a non--zero chiral condensate, such as QCD, we expect instead $\rho(\omega) = \rho(0) \neq 0$ in the chiral limit. Note that if we nonetheless perform a fit to the form of Eq.~(\ref{eq:rho}), this would give ``$\gamma_*$'' $= d-1$ in the limit $\lambda\rightarrow0$.

\section{Massive Schwinger Model}
The massive Schwinger model consists of $n_f$ Dirac fermions of mass $m$ which live in 1+1 dimensions, and interact through a U(1) gauge field. The euclidean Lagrangian is given by
\begin{equation}
\mathcal{L}(\overline{\psi},\psi,A_{\mu}) = \overline{\psi}_f(x)\left[\gamma_{\mu}(i\partial_{\mu} + gA_{\mu}(x)) + m\right]\psi_f(x) + \frac{1}{2}F_{\mu\nu}(x)F_{\mu\nu}(x),
\end{equation}
where the index $f$ runs from $1$ to $n_f$. For $n_f>1$ massless fermions, the spectrum of the theory consists of one massive boson of mass $\mu=\sqrt{n_f g^2/\pi}$, and $n_f-1$ massless bosons~\cite{Halpern:1975jc}.

For $n_f\leq 1$ the theory has a non--zero chiral condensate, whilst for $n_f \geq 2$ the condensate vanishes in the chiral limit. Assuming the fermion mass is light compared to the U(1) boson mass
($m \ll \mu$), and the volume is large compared to the U(1) boson mass ($\mu L \gg 1$), then the mass--dependence of this condensate is known in two limiting cases~\cite{Smilga:1992hx,Hetrick:1995wq}.
\begin{equation}
\left\langle \overline{\psi}\psi \right\rangle \propto \left\{
\begin{array}{lll}
m & \mathrm{for} & m L \ll 1 \\
m^{(n_f-1)/(n_f+1)} & \mathrm{for} & m L \gg 1
\end{array}
\right. 
\end{equation}

Comparing this to Eq.~(\ref{eq:condensate}), we see a very similar picture to the one we have for mCFTs; in the UV limit we have a free theory with $\gamma=0$, whilst in the IR limit $\gamma\rightarrow\gamma_*=1/n_f$. Hence for the $n_f=2$ case we consider here, $\gamma_*=0.5$.

\section{Simulation Details}
The implementation on the lattice used here follows Ref.~\cite{Gattringer:1997qc}, using the plaquette gauge action and unimproved Wilson--Dirac fermions. Configurations are generated using the Hybrid Montecarlo (HMC) algorithm, with trajectory length $\tau=1$, and the number of integration steps tuned to keep the acceptance rate in the range $60-90\%$. $\mathcal{O}(2000)$ configurations were generated for both $n_f=2$ and $n_f=0$, at $\beta=5,2,1$ on lattices of size $32^2$. A smaller number were generated at stronger bare coupling, $\beta=0.5,0.1$. Each configuration is separated by 20 HMC updates, and on each the lowest 300 eigenvalues are calculated using Chebyshev accelerated subspace iteration~\cite{DelDebbio:2005qa}. The Wilson--Dirac operator explicitly breaks chiral symmetry, so at each value of $\beta$, $\kappa\equiv1/(2m+4)$ is tuned to $\kappa\simeq\kappa_c$ (determined here as the value which minimises the lowest eigenvalue $a\Omega_0$). Some additional configurations were generated with larger masses, and on smaller lattices, to check for finite mass and finite volume effects.

\begin{figure}
  \centering
    \includegraphics[angle=0,width=13.0cm]{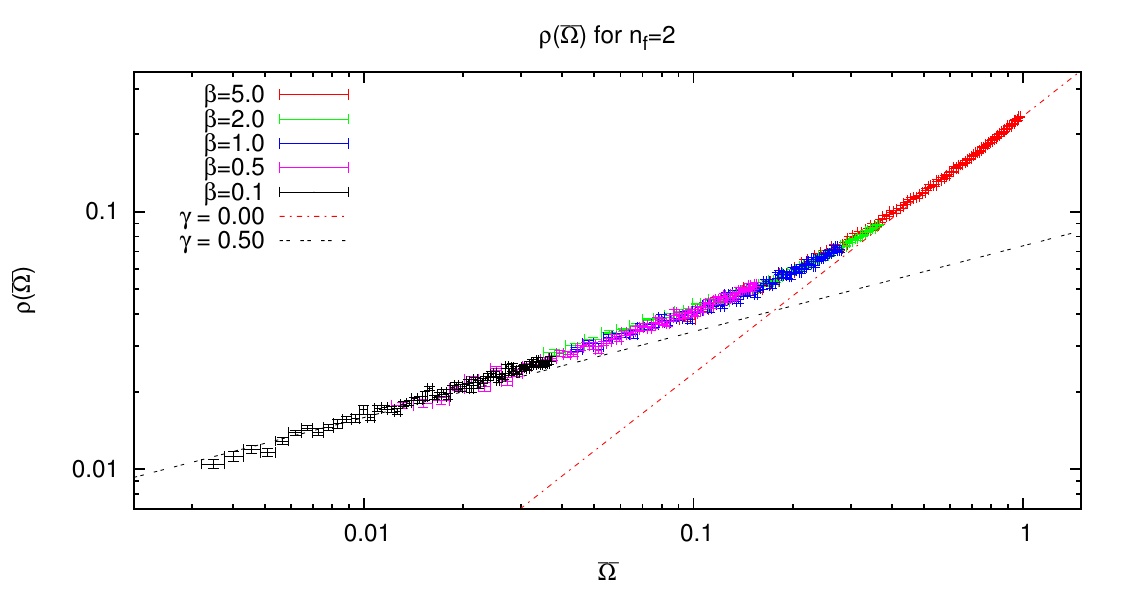}
  \caption{Log--log plot of eigenvalue density $\rho(\bar\Omega) \propto \bar\Omega^{\frac{1-\gamma}{1+\gamma}}$ versus $\bar\Omega$ for $n_f=2$ and various values of $\beta$. The slope changes from $\gamma\simeq0$ at large eigenvalues (red line shows the analytic prediction of the slope in the UV limit to guide the eye) to $\gamma\simeq \gamma_*=0.5$ at small eigenvalues (black line shows the analytic prediction of the slope in the IR limit). This demonstrates the importance of the fit range in determining $\gamma_*$.}
  \label{fig:rho_log}
\end{figure}

\section{Fit Function}
On the lattice we measure eigenvalues $\Omega^2$ of the massive hermitian Dirac operator $M=m^2 - \slashed{D}^2$. These are related to $\omega$ as $\omega = \sqrt{\Omega^2 - m^2}$. In general the lowest eigenvalues will be affected by finite volume and/or finite mass effects, and the highest eigenvalues will be affected by lattice artefacts, but we expect Eq.~(\ref{eq:rho}) to apply for eigenvalues in some intermediate range $\omega_{IR}<\omega<\omega_{UV}$. The strategy is then to determine $\gamma(\omega)$ for many values of $\omega$ within this range. We consider two ways to fit the data. The first is to fit the measured mode number directly, which assuming finite mass and finite volume effects are negligible can be done with the simple fit funtion
\begin{equation}
\label{eq:fitNU}
\nu(a\Omega) = A\left(a\Omega\right)^{\frac{2}{1+\gamma(a\Omega)}},
\end{equation}
where $A$ and $\gamma(a\Omega)$ are free fit parameters. Alternatively one can first construct the spectral density by binning the measured mode number, then do a fit to
\begin{equation}
\label{eq:fitRHO}
\rho(a\Omega) = B\left(a\Omega\right)^{\frac{2}{1+\gamma(a\Omega)}-1},
\end{equation}
where $B$ and $\gamma(a\Omega)$ are free fit parameters. 

In both cases we determine $\gamma(a\Omega)$ by fitting the data in the range $a\Omega \pm \Delta$, and both fits should give the same value for $\gamma_*$ at the IRFP. Using Eq.~(\ref{eq:fitNU}) gives more precise values for the fitted parameters because no binning of the data is required, but the price to pay is that finite volume effects below the fit range still affect the fitted parameters. Using Eq.~(\ref{eq:fitRHO}) avoids these finite volume effects, but increases the statistical errors, and there may also be a systematic dependence on the bin size.

To investigate a wide range of energy scales we use several different values of $\beta \equiv 1/g^2$, which correspond to different values of the lattice spacing $a$. The resulting $\gamma(a_{\beta}\Omega)$ can then be combined by rescaling the eigenvalues $a\Omega$ in terms of the lattice spacing at some fixed value of $\beta$~\cite{Cheng:2013eu}, in this case we use $a_{\beta=5.0}$. In the Schwinger model $a \propto 1/\sqrt{\beta}$, so we define the rescaled eigenvalues as
\begin{equation}
\label{eq:scale}
\overline{\Omega} \equiv a_{\beta}\Omega_{\beta} \left(\sqrt{\frac{\beta}{5.0}}\right)^{1+\gamma(a_{\beta}\Omega_{\beta})}.
\end{equation}

\section{Results $n_f=2$}

Fig.~\ref{fig:rho_log} shows the spectral density $\rho(\bar\Omega)$ as a function of $\bar\Omega$ on a log--log plot. The slope of the curve determines the value of $\gamma$, which runs from $\gamma\simeq0$ at large eigenvalues to $\gamma\simeq\gamma_*=0.5$ at low eigenvalues. For each value of $\beta$ the points have been multiplicatively scaled to make them lie on a single curve, this is only to make the plot clearer to the eye, and does not affect the slope of the curve. This plot shows the importance of considering a range of scales when determining $\gamma$ using this method.

\begin{figure}
  \centering
    \includegraphics[angle=0,width=7.5cm]{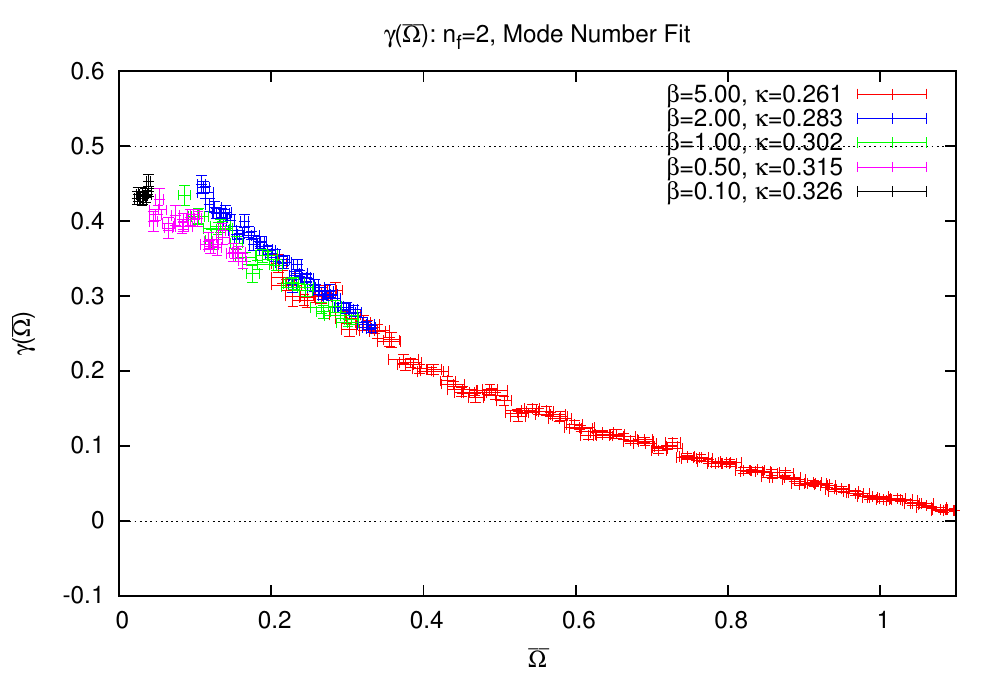}
    \includegraphics[angle=0,width=7.5cm]{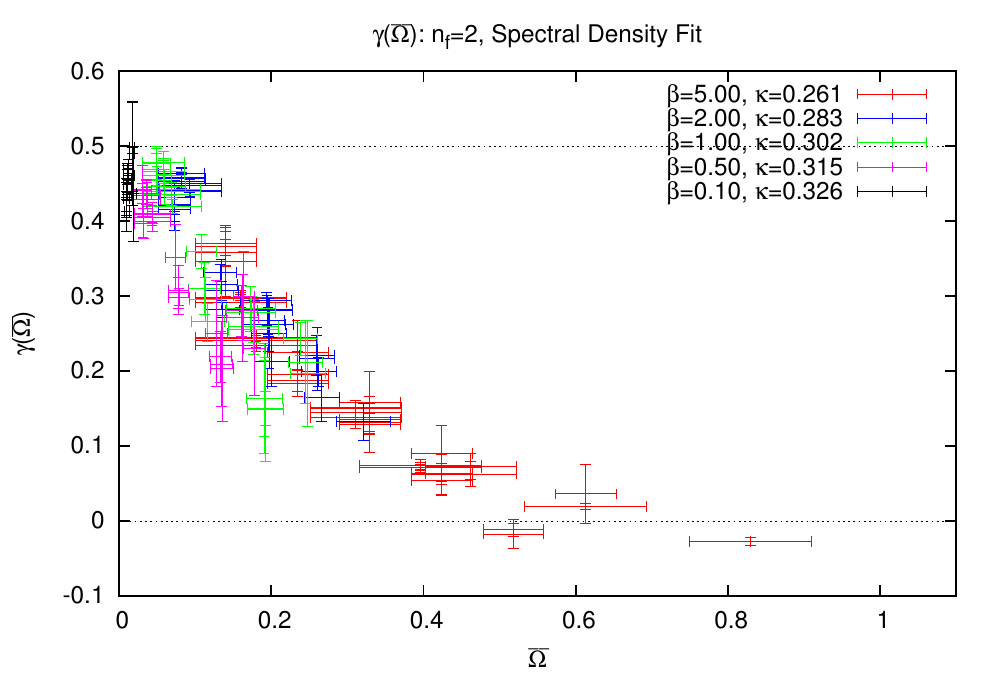}
  \caption{Mass anomalous dimension $\gamma$ vs eigenvalue for the $n_f=2$ theory. Left: mode number fit, right: spectral density fit. Different colours correspond to different values of $\beta$. The results appear to be consistent with the analytic prediction of $\gamma\rightarrow0$ in the UV (large eigenvalues), and $\gamma\rightarrow\gamma_*=0.5$ in the IR (small eigenvalues).}
  \label{fig:gamma_nf2_a}
\end{figure}

Fig.~\ref{fig:gamma_nf2_a} shows our determination of $\gamma(\overline{\Omega})$ for the $n_f=2$ theory. The left plot shows the values determined from fits of the mode number to Eq.~(\ref{eq:fitNU}), while in the right plot the spectral density, constructed by binning the mode number data, is fitted to Eq.~(\ref{eq:fitRHO}). Several bin sizes are used in the plot to check that there is no systematic dependence within errors on the choice of bin size. For each value of $\gamma$, the x error bar shows the range of eigenvalues used in the fit, and the y--error bar shows the statistical error of the fitted value.

At large eigenvalues, both fits give values of $\gamma$ consistent with zero, the analytic prediction in the UV limit. For intermediate eigenvalues the two fits give different values, but as one goes to smaller eigenvalues the two start to converge, as they should, and appear to be consistent with the analytic IR prediction of $\gamma_*=0.5$ at zero eigenvalue. Going to smaller eigenvalues would require a larger physical volume, either by using a larger lattice, or by increasing the lattice spacing. Note that although the fit ranges have been chosen such that finite volume and finite mass effects are negligible, no continuum limit has been taken so lattice artefacts remain.

\section{Results $n_f=0$}
\begin{figure}
  \centering
    \includegraphics[angle=0,width=7.5cm]{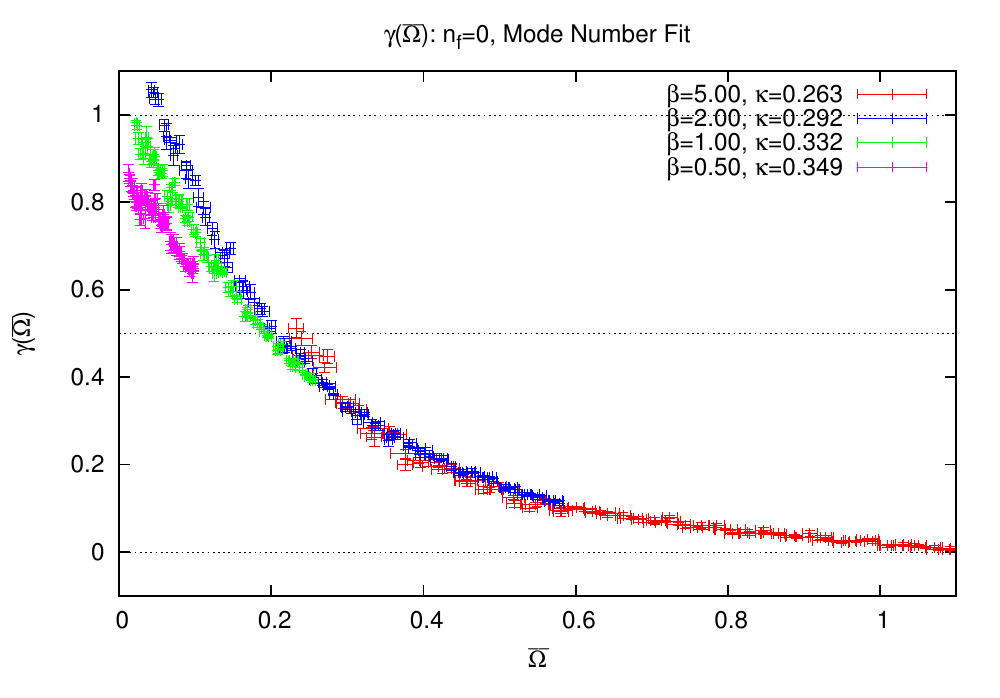}
    \includegraphics[angle=0,width=7.5cm]{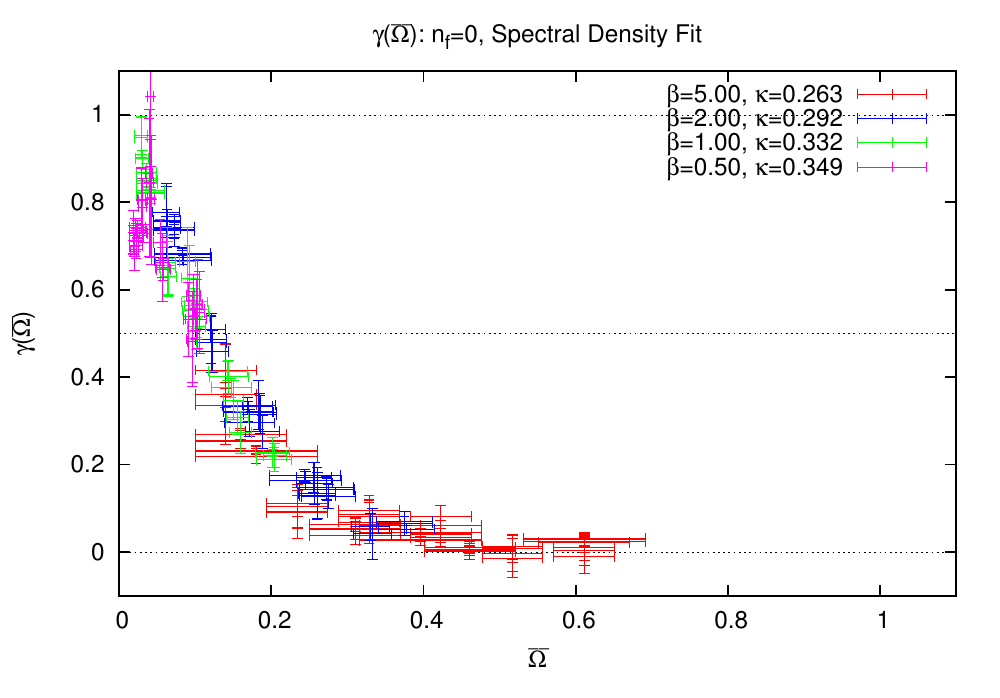}
  \caption{Mass anomalous dimension $\gamma$ vs eigenvalue for the $n_f=0$ theory. Left: mode number fit, right: spectral density fit. Different colours correspond to different values of $\beta$. In the IR limit (small eigenvalues), $\gamma=1$ corresponds to the expected chiral symmetry breaking.}
  \label{fig:gamma_nf02}
\end{figure}

We also apply the same method to the quenched theory with $n_f=0$. Fig.~\ref{fig:gamma_nf02} shows the resulting determination of $\gamma$ for this theory. Since this theory has a non--zero chiral condensate, we expect the mode number method to find ``$\gamma_*$''$\rightarrow d-1=1$ in the IR, which is consistent with what we find. The $n_f=0$ and $n_f=2$ theories clearly have different values of $\gamma$ for small eigenvalues, however at intermediate and larger eigenvalues there is little qualitative difference between these two theories, which shows the importance of considering a range of energy scales when trying to determine $\gamma_*$ for theories which may or may not have an IRFP.

\section{Conclusions}
For the $n_f=2$ massive Schwinger model, we were able to follow the running of the effective mass anomalous dimension over a wide range of energy scales using the Dirac mode number method, from $\gamma\simeq0$ in the UV to $\gamma\simeq\gamma_*$ in the IR. Our determination of $\gamma$ at the lowest measured energy scale is within $\sim10\%$ of the analytically known value in the IR limit. This accuracy could be improved by going to lower energy scales, either by using larger lattice volumes, or by using stronger values of the bare coupling to increase the lattice spacing, and performing a continuum extrapolation. A weakness of this method is that to avoid finite mass and finite volume effects, one is always forced to work at non--zero eigenvalue, and the appropriate form of the extrapolation to zero eigenvalue is not known, so that this distance from the IR limit remains as a systematic error that is difficult to quantify (unless one already knows the answer as is the case here).

We also applied the same method to the $n_f=0$ case, which has a non--zero chiral condensate. Whilst we found the expected behaviour at small eigenvalues, for intermediate eigenvalues the two theories have very similar values of $\gamma$, despite their vastly different IR behaviours, which underlines the importance of considering a range of scales when determining the mass anomalous dimension using this method.

Having confirmed that the Dirac mode number method reproduces the analytic predictions for this toy model, it would be interesting to use it to investigate the extrapolation to the IR (zero eigenvalue) limit, as well as the extrapolation to the continuum limit.

\section*{Acknowledgments}
We acknowledge use of the IFT clusters and CERN computing facilities. Thanks to Luigi Del Debbio, Margarita Garc\'{\i}a P\'erez, Antonio Gonz\'alez-Arroyo, Martin L\"uscher, Agostino Patella and Alberto Ramos for interesting and helpful discussions about this work.

\end{document}